\documentclass[letterpaper]{article} %

\pdfoutput=1

\usepackage{aaai24arxiv}  %
\usepackage{times}  %
\usepackage{helvet}  %
\usepackage{courier}  %
\usepackage[hyphens]{url}  %
\usepackage{graphicx} %
\usepackage{natbib}  %
\usepackage{caption} %
\usepackage{algorithm}
\usepackage{algorithmic}

\usepackage[inline]{enumitem}

\usepackage{newfloat}
\usepackage{listings}
\DeclareCaptionStyle{ruled}{labelfont=normalfont,labelsep=colon,strut=off} %
\floatstyle{ruled}
\newfloat{listing}{tb}{lst}{}
\floatname{listing}{Listing}
\title%
{%
Mind the Gap! %
Bridging Explainable Artificial Intelligence and Human Understanding with Luhmann's Functional Theory of Communication%
}

\author{
    Bernard Keenan\textsuperscript{\rm 1}\equalcontrib,
    Kacper Sokol\textsuperscript{\rm 2,3}\equalcontrib
}
\affiliations{
    \textsuperscript{\rm 1}School of Law, Birkbeck, University of London, United Kingdom\\
    \textsuperscript{\rm 2}Department of Computer Science, ETH Zurich, Switzerland\\
    \textsuperscript{\rm 3}ARC Centre of Excellence for Automated Decision-Making and Society, RMIT University, Australia\\%
    b.keenan@bbk.ac.uk, kacper.sokol@inf.ethz.ch
}

\newcommand{\cites}[1]{{\citeauthor{#1}}'s~(\citeyear{#1})}
\newcommand{\citeauthornoh}[1]{{\citeauthor{#1}}}
\newcommand{\citeauthorsnoh}[1]{{\citeauthor{#1}}'s}
\newcommand{\citeauthorssnoh}[1]{{\citeauthor{#1}}'}

\begin{document}

\maketitle

\begin{abstract}
Over the past decade explainable artificial intelligence has evolved from a predominantly technical discipline into a field that is deeply intertwined with social sciences. %
Insights such as human preference for contrastive -- more precisely, counterfactual -- explanations have played a major role in this transition, inspiring and guiding the research in computer science. %
Other observations, while equally important, have nevertheless received much less consideration. %
The desire of human explainees to communicate with artificial intelligence explainers through a dialogue-like interaction has been mostly neglected by the community. %
This poses many challenges for the effectiveness and widespread adoption of such technologies as delivering a single explanation optimised according to some predefined objectives may fail to engender understanding in its recipients and satisfy their unique needs given the diversity of human knowledge and intention. %
Using insights elaborated by Niklas \citeauthornoh{luhmann1992communication} and, more recently, Elena \citeauthornoh{esposito2022artificial} we apply social systems theory to highlight challenges in explainable artificial intelligence and offer a path forward, striving to reinvigorate the technical research in the direction of interactive and iterative explainers. %
Specifically, this paper demonstrates the potential of systems theoretical approaches to communication in elucidating and addressing the problems and limitations of human-centred explainable artificial intelligence. %
\end{abstract}

\section{Introduction\label{sec:intro}}

Research into explainable artificial intelligence (AI) is growing at a breakneck pace and becoming ever more complex. %
Researchers are exploring a number of diverse approaches, using different models of explainers across varying contexts, and devising a plethora of strategies for assessing the success of these distinct explainability tools, methods and media~\cite{guidotti2018survey}. %
When it comes to evaluating XAI, a whole new field of heterogeneous desiderata and criteria has emerged~\cite{sokol2020explainability,sokol2024does}. %
The question of whether or not a user can understand a data-driven automated decision-making (ADM) system is not simply a binary \emph{yes} or \emph{no} issue, nor it is reducible to the measurable psychological state of individuals~\cite{sokol2021explainability}. %
As such, the complexity of the problem of explainable AI (XAI) has grown.

It seems that the more we progress in XAI research, the more intricate the question and the broader the range of possible solutions become. %
We posit that this is because XAI is not simply a technological challenge that can be resolved with technical refinements alone. It additionally requires us to pay attention to the elementary building blocks of communication itself. But it appears that the closer we look at how satisfactory explanations are generated, communicated and accepted in natural communication, %
the more complex the problem becomes.  %

In this paper we draw on \emph{social systems theory}, as elaborated by Niklas \citeauthornoh{luhmann1992communication} and Elena \citeauthornoh{esposito2022artificial}, to offer a sociological account of explainable AI and its challenges. %
The theory provides a \emph{genetic} account of \textit{society as communication} that allows us to productively reframe the XAI field and highlight possible theoretical reasons for common challenges that researchers encounter.  %
Our work outlines a comprehensive selection of theoretical points, and then demonstrates them in an applied setting. %

The paper proceeds as follows. %
We first discuss the current state of XAI and the influence of insights from social sciences and psychology on (various technical developments in) this discipline (Section~\ref{sec:prelim}). %
We then introduce fundamental axioms of \cites{luhmann2012TOS1} theory of society, %
which allows us to highlight
why the problems of XAI are particularly resonant with systems theory's basic tenets (Section~\ref{sec:key_points}). %
Next, we show how the elements of communication pertain to problems commonly encountered in XAI research (Section~\ref{sec:selections}). %
With those epistemic concepts in place, we outline how the theory can be applied to XAI (Section~\ref{sec:systems}), and provide a concrete case study of XAI in healthcare, suggesting the questions that it may lead to (Section~\ref{sec:application}). We then conclude by revisiting our findings (Section~\ref{sec:conclusion}).  %

\section{Theorising Artificial Intelligence Explainability and Society\label{sec:prelim}}

Explainability has become one of the most important aspects of automated decision-making tools given their proliferation across different spheres of everyday life, especially in high stakes domains such as medicine, policing and justice~\cite{rudin2019stop}. %
Until recently, akin to the ``inmates running the asylum''~\cite{miller2017explainable}, many experts developing XAI systems relied predominantly on their personal intuition of what constitutes a ``good explanation'', paying little attention to findings from philosophy, social psychology or cognitive sciences~\cite{miller2019explanation}. %
Work on explainable AI has since striven to incorporate relevant insights on explanations and cognition to inform research into, and design of, data-driven algorithmic programmes and the interfaces that communicate their operation. %
These efforts aim to make the black box of AI explainable, in turn engendering trust in automated decisions and making them accountable where transparent actions are socially expected.%

While some insights from social sciences have since been widely adopted into technical designs by the XAI community, others remain largely neglected. %
The most prominent finding -- people's preference for contrastive explanations~\cite{miller2019explanation} -- has led the researchers to embrace \emph{counterfactuals}, which are statements of the form ``had a certain aspect of a situation changed like so, the outcome would have differed in this particular way''. %
Their flexibility, general comprehensibility and potential for compliance with legal frameworks~\cite{wachter2017counterfactual} as well as availability of generation procedures accounting for various human-centred properties~\cite{poyiadzi2020face,karimi2021algorithmic,romashov2022baycon} have further added to their appeal. %

In contrast, humans' propensity to \emph{interact with the explainer} through a dialogue-like interface -- another important insight from social sciences~\cite{miller2019explanation} -- is often overlooked in computer science literature~\cite{schneider2019personalized}. %
Since explanations are \emph{contextual}, i.e., the understanding of an explanation is always relative to the explainee's situatedness, beliefs, aims and assumptions about the explanation, presenting arbitrary insights into the functioning of data-driven systems is insufficient to generate trust, accountability and legal compliance, or to achieve other social ends. %
Explainability must thus be approached as an intrinsically social and interactive process of (natural language) \emph{communication}~\cite{finzel2021explanation,lakkaraju2022rethinking}.  %
Agency ought to be given to the explainees, who cannot be treated as passive receivers of information, but instead actively engage in bi-directional interaction with the explainer to facilitate their observation and selection of salient information leading to understanding~\cite{sokol2020explainability}. %

In view of these findings it may be difficult or even impossible to achieve meaningful and satisfactory algorithmic explainability without an appropriate \emph{communication framework} in place. %
For example, \emph{algorithmic recourse} -- which is a popular realisation of XAI that offers a sequence of actions guiding an explainee towards the desired outcome -- is only a partial realisation of this paradigm since human input is still limited or non-existent, making it a one-sided process~\cite{poyiadzi2020face,karimi2021algorithmic}. %
\emph{Formal communication schemas} and \emph{interaction protocols}, e.g., realised through a natural language dialogue, conversation or otherwise, have also been studied in this context, offering theoretical foundations, design patterns and frameworks for dynamic personalisation of explanations~\cite{walton2007dialogical,walton2011dialogue,walton2016dialogue,arioua2015formalizing,madumal2019grounded,schneider2019personalized}. %
These technical concepts aim to model and facilitate explanatory interactions between two intelligent agents, be they humans, machines or one of each, using explanatory and questioning (examination) dialogue modes -- e.g., implemented with formal argumentation~\cite{dung2009assumption} -- that allow the interlocutors to argue, dispute and challenge problematic aspects of (automated) decisions and their explanations. %
While insightful, these contributions are fairly abstract, predominantly theoretical and primarily grounded in technical research, yet difficult to implement, hence they have not yet been fully embraced by practical explainability tools with a few notable exceptions~\cite{kuzba2020would,malandri2023convxai}. %

Design, evaluation and deployment of XAI systems outside of academic research have proven to be challenging as well~\cite{bhatt2020explainable}. %
This is especially true for high stakes applications in which ante-hoc interpretability -- i.e., having a predictive model that is transparent by design -- is preferred~\cite{rudin2019stop,sokol2023reasonable}. %
Nonetheless, achieving this realisation of XAI requires more effort and domain knowledge than working with more universal and ubiquitous post-hoc approaches, where explainability is retrofitted into pre-existing data-driven systems by constructing a separate explanatory model. %
While producing insights that look like plausible explanations, the latter techniques often cannot guarantee their correctness, hence may be untrustworthy and misleading~\cite{sokol2022what,sokol2020towards}. %

Additionally, the properties expected of XAI systems span a broad range of \emph{sociotechnical} desiderata, many of which are context-dependent and some of which cannot be satisfied at the same time, requiring trade-offs~\cite{sokol2020explainability,sokol2021explainability}. %
The diversity of purposes to which end explainability is used and objectives had by distinct stakeholders involved in the process further complicate the creation, testing and deployment of XAI techniques in real life~\cite{bhatt2020explainable}. %
The lack of a universally agreed evaluation protocol also contributes to the problem~\cite{sokol2024does}. %
While generic frameworks exist, they cannot prescribe the exact procedure given the breadth of use cases and involvement of humans -- as explainees -- in the process, which itself may be iterative and interactive, thus further complicating explainability evaluation attempts.%

Even though realising explanatory communication in its full extent may be beyond our reach at the moment, its simplified version has been implemented for counterfactuals, allowing the explainees to customise and personalise their content via a rudimentary interaction schema~\cite{sokol2018glass,sokol2020one}. %
This limited form of user-driven information exchange within a fixed protocol is nonetheless already an improvement over simply making the \emph{interface} of an XAI technique interactive, which only offers an illusion of agency. %

More recently, the technical advancements in interactive XAI have been interpreted though the lens of a \emph{bi-directional, iterative social process}, and captured in a conceptual framework that allows us to better study explainability as a phenomenon that emerges from the operations of dynamic social systems~\cite{rohlfing2020explanation}. %
Since the explainees are not passive receivers and their mental states evolve throughout explanatory episodes, XAI should go beyond na{\"i}ve personalisation and model such encounters as interactions between two social agents: one providing cues that steer the explanations and the other responding with insights adapted to the explainee's current need for information. %
This process allows the explainer and explainee to \emph{co-construct understanding} by monitoring and scaffolding knowledge, which is built iteratively and kept contextually relevant at all times. %

Systems-based approaches have also contributed to theoretical analyses aimed at differentiating and elaborating the \emph{levels} of cognitive and social understanding required for data-driven models to operate in contingent and complex social and regulatory environments.  %
For example, \citet{dazely2021levels} have argued that designing tools for interaction with non-expert users necessitates developing \emph{conversational} models spanning multiple sense-making contexts, an approach they label \emph{Broad-XAI}. %
Their work %
adopts a constructivist approach to explanation, taking the cognitive process by which AI generates decisions as the model, and mapping different \emph{levels} of explanation to the \emph{human social process} involved in successful, trustworthy explanations. 
The technical realisation of an AI's \emph{cognitive model of its world} -- referred to as its \emph{merkwelt} -- then serves as a guide to XAI: %
\begin{description}
    \item[zero-order] 
explanations are reactive, they explain how an AI agent reacts to its perceived inputs; 
    \item[first-order] 
explanations concern the agent's disposition and motivations towards its environment, placing the reaction in the context of the agent's \emph{beliefs} or \emph{desires};
    \item[second-order] 
explanations are \emph{social}, in that they explain the decisions in relation to the agent's awareness of its own mental state and that of others;
and
    \item[n\textsuperscript{th}-order] %
explanations concern the \emph{cultural expectations} of other actors at a general level. 
\end{description}

The \emph{Broad-XAI} perspective is complementary to our systems theoretical methodology. %
Systems theory also entails a constructivist approach to reality -- it proposes an abstract explanatory model of communication in action. %
We thus argue that it provides a useful representation of the general social \emph{communicative environment}, offering a theoretical structure to the n\textsuperscript{th}-order level described by \citet{dazely2021levels} and elaborated further in Appendix~\ref{sec:closure_coupling}. %

We posit that systems theory offers a meta-theoretical framework that is useful for translating insights from across different social science approaches to XAI into technical requirements to be implemented by explainability tools. %
This can become especially helpful when ADM models operate in complex or high stakes environments, where the social importance of their explainers increases. %
In such cases, understanding
constructed and contested notions like \emph{meaning} and \emph{trust} must go beyond matters of technical design.  %
While the conceptual frameworks overviewed throughout this section fulfil many XAI desiderata~\cite{miller2019explanation,sokol2020explainability} and offer a blueprint for designing human-compatible interactive explainers, the literature overlooks the nature of communication itself. %
Here we address this gap by looking beyond the (technical) protocol responsible for the exchange of information. To this end, we employ the \emph{theory of social systems} and \emph{communication} introduced by \citet{luhmann1992communication,luhmann1995social} and elaborated by others~\cite{moeller2011radical}, most recently \citet{esposito2022artificial}. %

Despite the relevance of \cites{luhmann1995social} \emph{systems theory} to artificial intelligence explainability, the two concepts have only recently been connected, albeit superficially, to define explainability~\cite{sariyar2022medical}. %
The theory offers a response to the problem of the \emph{sociotechnical gap} separating social and technical systems, formulated by \citet{ackerman2003sociotechnicalgap} as the chasm between ``what we know we must support socially and what we can support technically''. %
As \citet{ehsan2023charting} have recently pointed out, XAI exemplifies this problem. 
Systems theory allows us to reframe this gap %
as the distinction between social and technical \emph{systems}.
Observed this way, the theory offers a useful set of methodological heuristics for mapping the \emph{social} environment in which ADM models and XAI tools operate,   %
which promises to unlock further progress in human-centred XAI. %

\section{Key Points of a Systemic Approach to XAI\label{sec:key_points}}

Before analysing the individual elements of communication (Section~\ref{sec:selections}) and applying systems theory to XAI (Section~\ref{sec:systems}) we first introduce a relevant selection of its fundamental axioms. %
A more in-depth overview of the theory -- specifically, the concepts of \emph{social systems} and \emph{communication} -- is available in Appendix~\ref{sec:communication_systems}. %

\paragraph{Society as Communication}
XAI is not simply a matter of imparting information or measuring human understanding, but a question of \textit{enabling communication}, which entails treating information and understanding as socially embedded phenomena. %
For systems theory, society is communication, and nothing but communication.
Furthermore, only communication communicates; individuals do not communicate.
Counter-intuitively, systems theory defines the individual as a \emph{psychic system} embodied in a \emph{living system}. %
Human consciousness is therefore \textit{coupled} to communication by the body's sensory and expressive capacities, but it cannot cross the systemic boundary.
Similarly, technology does not communicate.
Technical media like displays, speakers and input devices enable computers to be coupled to the social operations of communication, but not to communicate; computers and technical media are simply in communication's environment.
In \cites{luhmann1992communication} constructivist theory, communication is an emergent phenomena produced in social activity. The humans, texts, networks and devices that enable and shape communication are not \emph{in} it. %

\paragraph{Operational Closure}
Each social system adapts and orients itself to its environment, reproducing in each operation the distinction between system and environment for itself.
\citet{luhmann1992communication} insists on the \emph{operational closure} of systems, %
meaning no system communicates directly with any other system and information cannot directly \emph{transfer} from one system to another.
Instead, each system observes other systems as elements in its environment.
The environment, therefore, is not a given ontological reality, instead it is composed by the system itself; i.e., %
the environment is not the totality of what is ``really there'', but it is rather a cognitive construction of a system.
Like living organisms, systems must generate an image of the world for themselves by selecting observations and processing them.
In order to communicate about communication, we thus require a model for reducing complexity and dealing with self-reference.  %

\paragraph{Structural Coupling}
Although communicating systems are closed to one another, they nonetheless form \emph{structural couplings}, i.e., points at which different systems can mutually resonate or ``irritate'' one another in a regularised and predictable manner.
Structural couplings narrow the possibilities of each system's response to other systems and in so doing ``digitise analogue relations''.
They reduce the uncertainty of the environment by providing specific interpretative grids for one system to observe another, allowing precise information to be formed within both systems.
As mentioned above, the body's sensory organs link the social system of communication with the psychic systems of human beings as communication occurs~\cite{luhmann2012TOS1}. %
But structural couplings can be conceptual social artefacts too, for instance, a contract couples together legal communication about norms with economic communication about payments.
A coupling can be observed where both \emph{sides} of distinctly different modes of understanding are mutually \emph{irritating} one another.
The concepts of \emph{operational closure} and \emph{structural coupling} are further elaborated in Appendix~\ref{sec:closure_coupling}. %

\paragraph{Second-order Observation}
For systems theory, second-order observation -- which concept describes observing the observations of other observers, thus differentiating \textit{what} was observed from \textit{how} it was observed -- is the universal condition of modernity.
Put another way, there are no universal truths in modernity except second-order observations.
It does not matter whether the observer is human, machine, organisation or hybrid. %
Communication is therefore inherently marked by \textit{contingency}; %
each statement is always open to being observed otherwise.
In the context of explainability,
an ADM system generates observations via operations that cannot themselves be communicated.
The task of XAI is to bridge the gap between those technical operations and the social world of communication.
The XAI observes how the ADM model observed, and reports on it.
XAI research, in turn, observes the observations of human observers observing the observations of XAI.
Therefore, the task of XAI research is also second-order observation, as is this paper, which sits within a complex web of other observers' papers.

At all points, observations depend on \emph{reducing the complexity} of the world in order to make communication possible.
Additionally, rather than aiming at scientifically correct or perfect understandings of information, we instead need to consider how the contingency of XAI observations can be successfully operationalised by observing systems in their ongoing communication. %
Here, \emph{successful} communication does not require perfect understanding, nor it is about pure information transfer from the XAI system to the user, it only requires observing systems to be able to accept or reject new information with sufficient confidence in a given situation~\cite{sokol2020explainability}. %
Ultimately, the aim of XAI is to permit ADM systems to be deployed in complex social and organisational settings -- by making them comprehensible -- where data-driven predictions must be useful, trustworthy and reliable over time. %

\paragraph{Functional Differentiation}
Modern communication is functionally differentiated. %
Society is composed of differentiated \emph{systems} of communication, each of which relies on its own self-referential operations to produce stable sense over time.
Law and science are just two examples; in the first, one relies on legal procedures to produce normative decisions about what is legal and what is illegal, and in the second, one relies on experimental methodologies and assumptions to generate facts about what is true or false.
The differences are obvious but serve to demonstrate how differentiated systems of communication depend on different criteria of observation, which Luhmann described as \emph{coding} and \emph{programming}.

\emph{Systems} use their coding to connect one communicative operation to another over time, allowing meaning to be stabilised in differentiated domains, thus providing society with a background ``reality'' against which new information can be appraised and understood.
Systems make communication more likely to be successful by reducing the complexity of their environment and ``absorbing'' the contingency of their observations of new information. %
Put simply, when a highly unexpected event occurs, communication reacts in unpredictable and functionally differentiated ways; e.g., its legal implications are understood one way, just as its economic, political and scientific implications are each different again. %
The point is not to suggest that each of these differentiated systems of communication is isolated from one another, but to shift the perspective from individual humans to the systemic ways in which we understand the ever-changing social world. %

Notably, any individual person can observe events through the lens of any system.
Each system prompts different understandings and, at the second-order level, reduces the contingency of unexpected events by making them understandable in different dimensions.
Under this purview, therefore, ``understanding'' is not simply the mental understanding belonging to individual humans, but rather an emergent property of systems in general that unfolds inter-subjectively and knits communication together. %

This perspective showcases why XAI is a multifaceted problem with different approaches required in different contexts. %
The significance and expectations of a useful explanation change depending on the observing social system -- e.g., law, science, politics or education -- %
with each one containing sub-systemic examples. %
The radical constructivist approach of systems theory implies that all communication, including statements about causal effects, are contingent constructions produced by a system about itself and its environment -- constructions that could have been produced differently. 

The self-referential quality of social systems led \citeauthor{luhmann1992communication} to suggest that social systems are best understood as \emph{autopoietic}. %
\citeauthor{luhmann1992communication} adapted this biological concept to indicate that social systems reproduce themselves using only their own elements and that society is evolutionary, changing itself through the contingent emergence of adaptations, which are contingently picked up and stabilised by their own reproduction over time.

\paragraph{Autopoietic Dynamics}
Systems theory argues that society is evolutionary.
Communication systems that constantly reproduce themselves do so by communicating about their environments, which they construct for themselves.
Because the environment includes other communicating systems, each operating on its own terms, the environment of each system is in constant flux.
Communication is stimulated in its autopoietic reproduction by this inexhaustible dynamics of contingent environmental change and systemic adaptation.
The introduction of XAI as a means of allowing ADM tools to be used in communication thus entails multiple adaptive changes across a broad range of systems -- law, economy, politics, medicine, science and education are just some systems transforming themselves and their organisational forms. %
Therefore, just as XAI researchers adapt  %
technical systems to the needs of social systems, social systems are inevitably adaptive to XAI. %

\paragraph{Information Processing \& Organisations}%
Differentiated social systems \textit{expect} explanations for various reasons and provide distinct organisational and procedural modes of processing information. %
Ultimately, the point is to decide whether to accept or reject a given proposition in ongoing communication.
Even if a proposition is rejected, communication continues and learning can occur, either by correction or adaptation.
Traditionally, the provision of second-order explanations for information has relied on the work of human intermediaries in organisational settings, often specifically qualified to observe and evaluate evidence. %
Such observations are highly contextual and depend on the system in question as well as the importance of the statement in that system. %

Embedding XAI in social settings is not simply a question of making an ADM model understandable to a given human observer;
it is a question of evolving functional equivalence between XAI systems and the organisational systems and human expertise that previously ensured that decisions are trustworthy, reliable and generally accepted in communication~\cite{luhmann2018org}. %
Organisations learn by developing techniques for setting goals, generating information, deleting or retaining memory, distributing decision-making powers, conducting reviews, anticipating risks, carrying out plans, and reviewing the relationship of the organisation to its operational environment.
The accountability requirements of one observing system will be different to the requirements of others such as users, investigators, lawyers, insurers and the public -- or more precisely, the mass media that stand in for the public~\cite{dazely2021levels}.

The nature of organisational communication, in other words, allows it to adapt and absorb the uncertainty of navigating a complex environment~\cite{luhmann2018org}. %
Within this framework, organisational responsibility involves the tailoring of accountability procedures in ways that are irreducible to the states of the psychic systems of the organisation's individual members. %
A more in-depth discussion of the role of organisations in coping with the contingencies of communication -- especially in terms of decision-making accountability -- is provided in Appendix~\ref{sec:organisations}. %

\paragraph{Unpredictability}
As \citet{esposito2022artificial} explains, data-driven models are confronting us with information that is interesting precisely because it is not planned or available in advance -- unpredictable outputs that not even the programmers can directly explain. %
Artificial intelligence systems autonomously develop their procedures and identify patterns, later using them to generate contingent observations in response to our queries. %
The information output by AI does not precede the query, rather the algorithm generated it itself. %
In this context, XAI is tasked with enabling ADM models to function as effective \emph{interaction partners} in society's communication by allowing data to produce understanding that was not available in advance of the query. %
This is exactly the point of using data-driven algorithms -- they generate and thrive on the contingency of observations. %
Indeed, the very unpredictability of algorithmic decisions exacerbates the need for XAI even further since, as pointed out by \citet{esposito2022artificial}, ``we get information that often was not planned or available in advance and was unknown to the programmers themselves'', and, as noted above, it is offered as the basis for decisions about the future. %

\paragraph{Systems Theoretical Perspective on XAI}
With the fundamental axioms of systems theory in place, we can offer a new perspective on XAI by interpreting it through systems theoretical terms. %
Specifically, XAI research constitutes an emerging sub-system of science that responds to the problems of integrating the contingencies of ADM algorithms into communication. %
The challenge for AI explainability tools is to stabilise the coupling of algorithmic contingency to social systems. 
Where that happens successfully, communication has evolved,
potentially giving rise to 
new forms of \textit{communication media} as we discuss further in Appendix~\ref{sec:communication_media}.

\section{Elements of Communication:\\*Information, Utterance and Understanding\label{sec:selections}}%

In systems theory, communication has a tripartite structure: information, utterance and understanding~\cite{luhmann2012TOS1}. Communication is not the successful transmission of information from one system to another, from point $A$ to point $B$, but an internal process of an observing system; the system operates by producing information about its environment. Understanding is the observation of meaningful information differentiated from the utterance that carries it. Every observing system -- be it a psychical system of an individual human mind, a biological system like a neuron, or a social system of communication -- produces information for itself according to its own internal operations and criteria. The selections that it makes determine what is produced. 

This is not the same sense of \emph{communication} as used in \cites{shannon1949mathematical} information theory, %
which treats this process as a technical problem of successfully transferring quantifiable \emph{bits} of information on a channel from sender to receiver in the presence of noise~\cite{shannon1948mathematical,weaver1953recent,shannon1949mathematical}. %
Information theory deliberately excluded the question of what the transmitted information means. %
Systems theory, on the other hand, is concerned with the conditions and operations of meaningful communication, and presupposes that inter-systemic transfer is impossible.

Similarly, an XAI explainer must make \emph{selections} from among the complex operations carried out by the ADM system. In order to make the black box of an ADM algorithm transparent to an explainee, the explainer cannot simply transcribe the steps that were taken to reach the output value, as this would be too complex to be informative and would not be useful to the explainee~\cite{sokol2023reasonable}. %
Instead, the explainer must selectively reduce the complexity of the ADM system in a manner that counts as pertinent \emph{information} for the explainee~\cite{sokol2020towards}. In practice, neither ADM nor XAI systems are monolithic; they are complex and modular, and how they are arranged and programmed determines how they produce information about the world~\cite{sokol2022what,sokol2024does}. %

Once information has been selected according to the parameters of the ADM model and XAI system in play, the selected information must be arranged and presented in a form, or forms, that make it comprehensible to the explainee~\cite{sokol2024does}. %
This entails the selection of appropriate media that allow successful communication; for example, text, colour, numbers, graphs, charts, audio or voice. In systems theoretic terms, the rendering of information into mediated forms constitutes \emph{utterances}. 

Once information has been selected and uttered by the explainer, it can be observed by observers and \emph{understood} (or misunderstood). In systems theoretic terms, understanding takes place when an observer generates meaning by differentiating information from utterance. For instance, understanding occurs when an observer extracts meaningful information from the representation of data on a chart, table or statement~\cite{small2023helpful,xuan2023users}. 

When researching applied explainability, a common approach is to evaluate the accuracy and confidence of the \emph{understanding} developed in the minds of explainees (usually through user studies) after exposing them to the operation of an ADM model and the XAI framework built around it. %
Systems theory complicates this picture by highlighting the importance of the \emph{system} in which those tools are to operate. %
Individuals act as coupling points within observing systems and must communicate through them.
For example, a doctor observes information on a screen of a medical device provided according to the \emph{protocols and priorities} of the medical system. %
The systemic aim of such guiding principles is to ensure that the decisions expected of a doctor are made within the communication of the medical system and its organisational sub-systems. %
How a doctor understands such information -- e.g., a prediction output by an ADM model -- about a given patient and how he or she operationalises it are highly dependent on the protocols of the situation. %

The point is that the correctness and confidence of the information transferred from an ADM model to the human mind is not alone sufficient to successfully operationalise this newly gained knowledge. %
Systemically effective understanding emerges according to the codes and programmes of the observing system as it observes and processes the information produced by the ADM model and its explainer.
Therefore, XAI needs to incorporate these systemic elements, recognising that such technological tools always operate in the environment of a communicating social system. %

\section{Systems Theory Applied to XAI}\label{sec:systems}

With the foundations of systems theory in place, %
we can now map the three basic elements of \cites{luhmann1992communication} concept of communication to artificial intelligence explainability as follows.%
\begin{description}
  \item[Information] becomes the salient (numerical) insights about an ADM model's operation generated by an explainer.%
  \item[Utterance] is embodied by the social construct used to convey the \emph{information} to a user, e.g., explanation type, explanatory artefact as well as communication medium and protocol.%
  \item[(Mis)understanding] captures successful \emph{sense construction} of the ADM model in the explainee's mind and its relationship to the social systems with which he or she is observing. The explainee's mental state is, of course, measurable only insofar as it is made observable via communication, and successful to the extent that the communication can successfully incorporate the explanation as salient in order to use it in further communications, or reject it as inadequate for specific reasons. %
\end{description}

As computers do not \emph{understand} their own operations and do not know what they do not know -- they are indifferent to the meaning of information they process and the objective of the processing algorithms -- an elementary algorithmic explainer must be designed around the first two variables: \emph{information} and \emph{utterance}. %
The explainer must first \emph{select} salient information from the operation of the ADM model in question, e.g., a decision that it outputs; these selections are contingent on the parameters and configuration of the explainer. %
Notably, the explainer is, in systems theoretical terms, a second-order digital observer of the underlying ADM model, which is in its environment (or \emph{merkwelt}). %
How a given explainer selects information from its constructed environment is a technical question, as is how it internally processes the selected observations to produce an explanatory insight~\cite{sokol2020towards,sokol2024does}. %

Information selection can either be based on a predefined set of criteria, e.g., embedded in an optimisation objective that is formalised by the XAI designers~\cite{romashov2022baycon}, or it can be delegated to the explainees, e.g., through a user interface or an interactive explanatory process~\cite{sokol2018glass,sokol2020one}; a mixture of the two is also possible. %
With respect to how an XAI tool internally processes information, since its encoding needed to achieve high predictive power may be unintelligible to the envisaged audience, the explainer may need to \emph{translate} it into a more suitable format, e.g., a domain-specific interpretable representation~\cite{sokol2020towards}. %
Next, the explainer must be designed to select appropriate artefacts to present its findings. %
There are multiple possibilities, with XAI researchers exploring how diverse formats such as numerical, visual, graphical and textual media differ in their capacity to engender understanding~\cite{small2023helpful,xuan2023users}, %
as well as on a more fundamental level how choosing contrastive or associative explanation forms impacts explainability~\cite{celar2023people}. %

Ultimately, understanding will depend upon the observer.
The \emph{what} of information is differentiated from the \emph{how} of presentation to produce understanding, including the possibility that more information is needed or a different form of utterance is required~\cite{sokol2024does}.
While currently such situations often cannot be resolved by the XAI tool itself, they can, and should, be anticipated by its creators. %
The design task is thus finding combinations of \emph{selections} that make successful communication more probable than not, and to design \emph{conversational} systems that allow communication to continue in contexts and domains where particular characteristics of explanations count as informative. %
Each selection will always be open to second-order observation.
The observer may ask ``Why this and not that?'' -- contrastive or counterfactual explanations -- or ``What goal was being sought?'' -- a more complex functionalist explanation -- and it will frequently be important to maintain the possibility of asking such questions %
for some time into the future after the decision was made, depending on the observing system in question, as different queries may become salient at different times~\cite{corti2024moving}. %

If the explainer fails, then communication ceases; the social system rejects the information as it is understood. %
As discussed above, a key distinction between \emph{artificial} and human intelligence is that AI cannot engage in understanding -- %
the explainer can only \emph{simulate} communication, it cannot engage in it. %
Options and limitations for the continuation of communication must therefore be designed and incorporated into the explainer in advance and this must allow for different levels of observation. %
Notably, there is no one \emph{true} explanation in each case but rather a range of possible selections of information and utterance~\cite{sokol2023navigating}.
Explanations are second-order observations of indications but are at the same time second-order forms in a particular medium. %

This perspective agrees with experimental findings that report study subjects feeling that ``there is not enough information'' provided by an explainer to allow them to trust that the explanation they received is a fair, non-arbitrary insight into the functioning of a predictive algorithm~\cite{schoeffer2022there}. %
Unless the user of an explainer is also a programmer with access to its coding and relevant expertise, there is little scope for recourse and reconciliation. %
Communication with an explainer is one-sided since neither the predictive model nor the explainer can ``change its mind'' because either lacks agency, ingenuity and creativity to adapt to human users. %
This leads us to agree with \cites{esposito2022artificial} insight that the problem as a whole is not one of dealing with an artificial \emph{intelligence} akin to a human interlocutor but rather a problem of designing for \emph{artificial communication}.  %

An explanation, whether accurate or not, will change the state of the system or the \emph{mental model} of the user when it is included.
This in turn gives rise to other observations, whether directly related to the ADM model or other aspects of the explainee's worldview. %
Such artefacts and externalities cannot be anticipated in their totality as it is not possible to include everyone's views in advance or foresee spontaneous thoughts and ideas that an explanation may trigger. %
A na{\"i}ve example of such a situation is the explainee incorrectly generalising an explanation to other, unrelated outputs of the predictive model in question~\cite{small2023helpful,xuan2023users}. %
But systems thinking can provide structure and depth to the XAI design task that go past the psychic response of any given individual. %
Such a perspective allows us to %
move beyond the figure of the user and study decisions in their social systemic sense-making contexts. %
These considerations may include %
organisational forms, in which case the design of XAI systems must account for organisational procedures (see Appendix~\ref{sec:organisations} for more details). %

Additionally, the anticipated functions of the explained ADM model should be carefully examined to understand how they emerged as necessary elements in relation to the dominant elements %
of the social system in question, how the information necessary for such decisions has been selected in the past, and the forms in which it has been expressed and made available for second-order review.
Notably, this perspective agrees
with the aforementioned \emph{Broad-XAI} approach~\cite{dazely2021levels}, which encapsulates
the need to investigate the social and cultural expectations in an agent's environment,
with a good explanation using them as its elements. %
Such an investigation may allow XAI designers to form a systemic understanding of how a particular decision-making point emerged socially as a problem to be solved, rather than taken for granted. %
Similarly, a systemic approach would ask how decisions are currently handled, the temporal frame in which they are processed and decided, and how they are assessed, accepted or rejected.

Furthermore, the design process should identify the interested parties, stakeholders and potentially affected groups, organisations and individuals, along with the relevant legal, political and normative considerations that inflect on it. %
Some decision contexts will simply require a greater variety of available explanations than others. 
From this perspective, judging the effectiveness of XAI is a problem of selective understanding, hence a recursive social problem -- %
rather than being a self-contained technical question, it has recursive and iterative dimensions.
As technology is adapted to the contingencies of communication, communication in turn adapts to accommodate the contingencies of technology.
Social systems thus adapt to changes in the structure of communication. 

While not explicitly evoked, \cites{luhmann1992communication} functional theory of communication can be seen throughout different areas of (technical) research dealing with explainability of data-driven models, especially so in the field of human--computer interaction, which aims to bridge the aforementioned sociotechnical gap inherent to XAI. %
For example, scholars in this domain assume a fixed explanation type -- such as a counterfactual statement delivered in text -- and tweak various aspects of its presentation and the underlying (explanatory) user interface in an attempt to make it -- or in the systems theoretical terms \emph{the act of communication} -- more likely to be accepted by a selected group of explainees~\cite{du2019eventaction,hadash2022improving,bove2023investigating}. %
Within this context, our contributions reinforce the, often neglected, purview that the explainee is not only a recipient of an explanation but also a \emph{communication partner}. %

The consequences of disregarding this facet of XAI %
can be seen in explanations output by state-of-the-art explainers deployed in real-life applications being ignored or rejected, which may simply be an overlooked symptom of inadequate communication~\cite{sivaraman2023ignore}. %
This broadened perspective accounts for the operational context of XAI systems -- including the stakeholders, purpose of explainability and the like -- in a more holistic way, thus allows us to better understand how ADM models and explainers thereof fit within society and its spheres. %
The connection between XAI and \cites{luhmann1992communication} functional theory of communication outlined in this paper can therefore help us to better choose \emph{information} and \emph{utterance} that are appropriate and sufficient to make sense of what is happening inside an algorithmic black box in view of the world that surrounds it~\cite{ehsan2023charting}, thus bridging the gap between automated decision-making and human \emph{understanding}, as well as reconciling the technical and social aspects of XAI research. %

\section{Case Study of XAI in Healthcare\label{sec:application}}%

Finding a suitable setting for the deployment of ADM models involves identifying current areas of practice where existing processes can be enhanced by data-driven analysis in a measurable way. %
Healthcare is a unique domain in this regard as even minute improvements in medical practice reap large societal and economical rewards~\cite{johnson2023potential}. %
To date, ADM tools have enhanced detection of diabetic retinopathy, classification of skin cancer and metastases from breast cancer~\cite{gulshan2016development,esteva2017dermatologist,golden2017deep};
nonetheless, many such models remain black boxes despite their adoption and the benefit thereof. %
The challenge for XAI in this context is thus to enable better integration of data-driven predictive models into the social system of healthcare. %

Existing medical workflows and decision-making practices are well established~\cite{chang2020intelligence}. %
The \emph{autopoiesis} of the healthcare system is concerned with the management of patient trajectories as well as investigating, monitoring and curing medical conditions~\cite{berg1999patient}. %
Patient treatment follows protocols established to assess and anticipate outcomes, and organisations -- here, healthcare institutions -- have workflows designed to implement treatment plans while absorbing the contingencies of unknown variables and unexpected changes. %

Exactly this complexity makes the adoption of new technologies highly challenging. %
This can be seen in the \emph{translational barrier} -- a chasm between technical solutions and clinical applications -- that is prevalent in the AI for healthcare research~\cite{wiens2019no,kanjilal2020decision,moehring2021development,adams2022prospective,markowetz2024all}. %
Notably, it can be linked to the sociotechnical gap at the interface of society and technology that XAI strives to address. %
This phenomenon leads to healthcare remaining one of the least digitised social systems, with many open challenges despite significant (technological) progress in the recent decades~\cite{berg1999patient,capobianco2019data,spatharou2020transforming}. %
While there are plenty of AI models claiming state-of-the-art performance on selected tasks, these tools rarely ever become integrated into real-life medical workflows or even tested in a clinical setting, which impedes their adoption as %
such (non-clinical) evaluation results are unlikely to directly translate to clinical efficacy~\cite{topol2019high,wardi2023bringing}. %

While technical challenges unequivocally contribute to the translational barrier, the inherent incompatibility of ADM models with individual, organisational and social aspects of decision-making workflows -- especially in high stakes domains such as healthcare -- is also a contributing factor~\cite{simkute2022xai,anjum2023conversation,wosny2023experience}. %
When building data-driven predictive tools the conception of their role (in deployment) tends to be assumed too narrow~\cite{mueller2019explanation,topol2019high,akata2020research,ferrario2023experts}. %
These systems are usually tasked with streamlining or automating (decision-making) tasks thus far handled by humans, for which they can display impressive evaluation results, often leading to a conclusion that they achieve performance higher than that of domain experts with decades of professional experience. %

Once equipped with human-centred interpretability or explainability, it is argued, such data-driven predictive tools could become worthy replacements of fallible humans. %
But such a view of automation appears too limited to be successful in real life~\cite{munn2022automation}. %
Decision-making is just one, albeit acutely observed, step in the complex network of tasks and duties held by every stakeholder, %
who may as well have multiple, possibly inter-dependent, roles with different responsibilities across a number of organisations~\cite{siddarth2021ai}. %
Replacing just one, artificially carved out, component in this network with data-driven automation may disrupt all the other well established processes, causing apprehension towards automation and leading to it being widely ignored~\cite{sivaraman2023ignore,tricco2023implemented}.
In systems theoretical terms, there is simply too much contingency in play, and too little trust in the reliability of AI decisions. %

Replacing humans with automation is also problematic because the attribution of responsibility for algorithmic decisions remains unclear~\cite{van2015epistemological}. %
Whether AI is used for full automation, where ADM tools make and execute a decision with only human supervision, or it simply supports human decisions, e.g., by offering a suggestion or justification of a particular data-driven action for the human to consider, approve/reject and implement, when either workflow brings unintended consequences assigning the blame is non-trivial. %
This is of particular importance in high stakes domains such as healthcare, where mistakes can have dire consequences and tend to be mediated through the legal system. %
Law places demands for regulatory compliance on providers and, where things go wrong, assigns liability. 

Ethnographic observation of the use of ADM tools in healthcare practice has demonstrated how second-order observation occurs.
For instance, when confronted with a prediction made by the \emph{Sepsis Watch} model, doctors reported wanting to ``interpret the model'' in the sense of ``understanding the causation''. Yet as the model was ``totally uninterpretable'', the development focused on demonstrating its efficacy within pre-existing social relations~\cite{elish2018stakes}. %
In a similar vein we suggest not to introduce ADM tools for \emph{decision} support, but rather to build systems that aid human \emph{reasoning}~\cite{van2021clinical}. %
This distinction is especially pertinent given %
that AI decision support is often understood as implementing an ADM system that mimics a selected task that thus far has been in the domain of human agents, and presenting its conclusions, possibly along with algorithmic explanations, to people. %
But as we have discussed earlier, this view of an action being self-contained and atomic completely disregards the nature of the decision-making process and its broader systemic situatedness.
Explanations that merely address the accuracy of an AI prediction and how it was technically produced are simply insufficient to meet the demands of complex social systems. 

Indeed, XAI can actively hinder the use of ADM tools~\cite{kaur2024interpretability}. %
Providing an AI prediction and justifying it with insights derived from an algorithmic explainer may lead to disengagement, under-utilisation of human expertise as well as automation bias, or in the worst case ignoring ADM tools altogether~\cite{byrne2023good,miller2023explainable}. %
Instead, we should strive to create data-driven systems that the users can interact with to better understand the underlying decision-making environment (e.g., through exploration), appreciate the uncertainty inherent to this domain, %
hypothesise about various actions and their consequences, and the like. %
This approach can additionally help humans to identify important, previously overlooked, observations and learn~\cite{crandall1991guide,crandall1993critical,hogarth2001educating}. %

Operating within the reasoning support framework also keeps the responsibility for the decisions with humans in the formal positions they occupy in systemic organisations. %
In such a scenario an ADM model is simply a tool, and as long as it behaves predictably -- i.e., it is robust and trustworthy, which may, for example, be guaranteed via adequate certification processes -- the insights developed by interacting with it can be safely incorporated into human decision-making. %
In this case, the users bear the ultimate responsibility for their actions, the accountability of which is achieved through their ability to justify individual decisions by supporting them with information or evidence provided by the ADM tool as well as their own knowledge, expertise and experience. %

This schema has already been incorporated into clinical workflows, albeit with systems that are primarily based on deterministic technology and that passed medical certification, for example, a machine to analyse blood samples. %
Doctors thus do not normally question the validity of blood test results since the underlying mechanisms are guaranteed to produce reliable outputs and the contingencies have been shifted to other entities, e.g., the device manufacturing company, the relevant government certification body, the technician responsible for regular inspections of the device, and the like. %
This operational framework allows doctors to simply integrate the resulting biomarkers into their cognitive process spanning activities such as exploration, analysis, hypothesis formation and (in)validation, decision-making, and action execution. %
As systems theory holds, organisational forms evolve as communication adopts ways of reducing the complexity and risk of its contingent environment.

Therefore, we should not just strive to automate a task but also to understand the (organisational) structure of the selected decision-making workflow, and ensure that the replacement ADM tool is designed with all the roles for all the stakeholders in mind, adhering to their expectations. %
Once accounted for, making data-driven models interpretable or explainable may still be insufficient to enable their integration with humans, and viewing this process through the lens of communication outlined by systems theory can offer useful guidance. %
In particular, even if an ADM tool is certified as reliable, this is no guarantee of perfect information transfer or systemic trust. %
Second-order observation means that unpredictable perspectives may be brought to bear, or contingencies emerge that cannot be designed for in advance. %
The complexity of the setting and the risks that ADM tools may add to the organisational operations therefore need to be carefully assessed. %

Introducing data-driven automation without taking the aforementioned steps can inadvertently complicate matters that such systems are intended to simplify. %
In summary, a good starting point is to: %
\begin{enumerate*}[label=(\roman*)]
\item
design %
for the programmes that currently exist, %
\item
identify where an ADM model might be usefully deployed, %
\item
tailor the explanations to the second-order questioning that communication generates, and %
\item
ensure that the distribution of risk and responsibility within the organisational structure (e.g., of the medical system) is maintained and supported but not overturned by AI tools. %
\end{enumerate*}
By viewing these interactions through the lens of communication we can better design artificial intelligence and its interpretability or explainability. %

\section{Conclusion\label{sec:conclusion}}%

While throughout this paper we have demonstrated the usefulness of systems theory for XAI, as it stands it remains just that, a highly abstract theory. %
According to \citet{moeller2011radical}, \citeauthorsnoh{luhmann1992communication}~(\citeyear{luhmann1992communication}) theory is a ``fourth insult'' to humanism. %
\citet{freud1920general} identified three insults to anthropocentrism; \citeauthorssnoh{copernicus1543revolutions}~(\citeyear{copernicus1543revolutions}) astronomy, \citeauthorsnoh{darwin1859on}~(\citeyear{darwin1859on}) evolutionary biology and \citeauthorsnoh{freud1920general}~(\citeyear{freud1920general}) psychoanalytic theories respectively challenged the centrality of humanity in relation to the cosmos, to creation and to the mind. %
\citeauthorsnoh{moeller2011radical} point is that \citeauthornoh{luhmann1992communication} similarly de-centres humanity in society.
Humans are in the environment of communication, which cannot be steered and has no unifying point of control.
Rationality is contingent on systemic conditions and cannot be universalised or unified.
The evolution of society is contingent on society's own operations and therefore cannot be directly controlled or planned for. 

We are at an inflexion point in the evolution of communication marked by an acceleration in the shift of information processing from psychic to computer systems.
Algorithmic tools analyse information without relying on understanding. %
The challenge of XAI is therefore critically important as it forms the structural coupling point at which meaning and understanding emerge in relation to artificial intelligence agents.
Echoing \citet{esposito2022artificial}, we suggest that XAI is key to the development of successful artificial communication.
We could even think of XAI research as an emergent social system that is evolving its own techniques and programmes at the interface of algorithmic systems and communication. %
Inevitably, XAI research will attract criticism and critique.
A successful explainer tool can be observed as enhancing trust and, at the same time, criticised for increasing automation bias~\cite{schoeffer2022there}. %
Such differences in perspective nonetheless cannot be definitively resolved scientifically as they embody second-order observation. %

In this paper,
we proposed an epistemic shift from minds to systems, from interpretations to communication. %
The task for researchers is to establish new structures that can be selected for use or indeed, in situations where understanding is legally or politically essential to the acceptability of a decision, for prohibition -- a point forcefully made by \citet{rudin2019stop}.
But we must at the same time be attuned to the limits.
Computers are not autopoietic systems and do not perceive themselves;
they do not know what they do not know.
One outcome of the deployment of ADM and XAI systems in society may be that firmer lines are drawn around values, decisions and organisational capacities that must be reserved to the psychic operations of human beings.
Paradoxically, a theory that excludes us, human beings, from social operations might help us see how to better defend ourselves.
We can specify the importance of human decision-making, meaning-creation and the possibility of accountability in differentiated domains. %
After all, ``if the outcome of a traditional machine becomes unpredictable, we do not think that it is creative or original -- we think that it is broken'' as observed by \citet{von1985cibernetica} and reported by \citet{esposito2022artificial}. %

\section*{Acknowledgements}
This research was conducted by the ARC Centre of Excellence for Automated Decision-Making and Society (project number CE200100005), funded by the Australian Government through the Australian Research Council. %
The work was done in part while the authors were visiting the Simons Institute for the Theory of Computing at UC Berkeley.%

\bibliography{manuscript}

\cleardoublepage
\newpage

\appendix

\section{Differentiated Social Systems and Communication\label{sec:communication_systems}}

Social systems theory is a model of society premised not on the agency of individuals or institutions, but on the conditions for the reproduction of communication. For \citet{luhmann1992communication}, society is constructed exclusively of communication, and only communication communicates. %
This is not the same sense of ``communication'' as used in \cites{shannon1949mathematical} information theory, %
which treats communication as a technical problem of successfully transferring quantifiable ``bits'' of information on a channel from sender to receiver in the presence of noise~\cite{shannon1948mathematical,weaver1953recent,shannon1949mathematical}. %
Information theory deliberately excluded the question of what the transmitted information means. %
Systems theory is instead concerned with the conditions and operations of meaningful communication. According to the theory, society is a complex of self-referential autonomous systems of communication. Communication is a self-organising process of differentiation that, independently of any central control, evolves and differentiates codes and structural processes, and does so using only its own processes. Systems make meaning possible by reducing the complexity of the world in order to communicate about it, which in turn makes society more complex as it communicates about itself and its environment. 

The starting point is the distinction between system and environment. A system is constituted by the distinction between itself and its environment. The environment is always more complex than the system, which responds to the complexity by ordering itself so as to adapt and interpret the environment by generating information about it. Social systems are operationally closed but cognitively open to an environment that is, for the system, a correlate of the system's own operations. The system indicates its environment by excluding it from the system, drawing a distinction between the world and itself. Paradoxically, this means that the ``reality'' of the distinction between system and environment cannot be grasped as a totality, as to indicate the distinction itself depends on drawing another distinction. 

Every system observes and is made observable by the difference between itself and its environment -- a distinction that is made within the system itself. Systems adapt and orient themselves to their environment,  reproducing in each operation the distinction between system and environment. \citet{luhmann1992communication} insists on the operational closure of systems; %
no system communicates directly with any other system and information cannot ``transfer'' from one system to another. Rather, each system observes other systems as elements in its environment, and responds to its observations of other systems only on its own terms. The environment is not a given ontological reality, rather it is open-ended and mutable horizon of experience composed by the observing system in question. In other words, the environment is not the totality of what is ``really there'', it is a cognitive construction that each system produces for itself. This way, order is generated from noise. Like living organisms, systems must discover the world by and for themselves -- they produce information internally by selecting observations and processing them. 

When applied to social systems, there is no way to communicate about the distinction between society and the world without relying on communication. We must observe communication using communication. Hence totalising terms like ``society'' and ``communication'' serve only as references -- symbolic terms that have particular valences when deployed in the ongoing communication of self-referential social systems. To understand and communicate about communication, we must rely on communication. Systems theory is in this way premised on paradoxes of observation, and this, as we suggested above, makes it a uniquely interesting way for charting the paradoxical problems of the \emph{sociotechnical gap}~\cite{ackerman2003sociotechnicalgap}.

As explained in Section~\ref{sec:selections}, the theory defines communication as a tripartite process of differentiation. Each communicative operation is composed of three distinct \emph{selections} from a horizon of possibility. In each communication, \textbf{understanding}
is the key moment; it occurs through the drawing of a distinction between selected \textbf{information} and \textbf{utterance}. Understanding -- the making of meaning -- is observer-dependent and arises in the decoding of \emph{what} is communicated from \emph{how} it is communicated. With each selection, an element is actualised from a horizon of potentiality by means of a distinction. What is indicated is distinguished from what is not indicated. This is why each communicative operation reinstates the boundary between an observing system and its environment. One side of the distinction is marked, the other side is excluded from indication. 

The inherent contingency of each selection can, in turn, be observed via further second-order operations, which follow in time and can indicate both the previous indication and its \emph{unmarked side} -- making an indication of what was not selected. In this way, the first distinction re-enters the system. The system refers to its own previous operation and in so doing performs a new operation -- a new instantiation of the distinction between system and environment. Thus each second-order observation entails the use of a further selection by means of a further distinction, the contingency of which can only be observed by means of a further second-order distinction, and so on. 

Therefore, there is no ultimate observation point; no privileged objective measure that is not made within a system. Each observation requires making a selection and thus produces a blind spot that can itself only be observed by way of further observations. If one continually utilises second-order observation to interrogate a previous observation, the system enters infinite regress. For this reason, systems have no ``origin'' and no prime causal factor. All attempts to identify a simple explanatory cause, or a foundation point for society's operations, only results in further operations, further second-order observations of observations, which may agree or disagree. Systems theory therefore rejects any claims to objective causal explanations for society as these are precluded by its own autological structure. Causality can only be observed within the operations of social systems: legal causality, scientific causality or philosophical causality. These are, of course, different forms of causality that each depend on system-specific axioms and procedures. Moreover, in today's complex society, they are all contingent on the specifics of the case in question.  

This self-referential autonomy of social systems led \citeauthor{luhmann1992communication} to suggest that social systems are best understood as \emph{autopoietic}. This term, derived from studies of the reproduction of living cellular systems, indicates that social systems reproduce themselves using only their own elements and that society is evolutionary, meaning that it changes through the contingent emergence of changing adaptations, which are contingently picked up and stabilised by their own reproduction over time. 

If we are all ultimately trapped in the labyrinth of communication, then we must be careful in how we theorise communication. Communication is concerned with the moment of understanding emergent in the observing system, but does not depend on \emph{accurate} understanding. %
Rather, a ``successful'' communication is merely one that is accepted as the basis for further communication. To this extent, \emph{all} communication can be regarded as productive misunderstanding -- it is misunderstanding, clarification, disagreement and self-correction that keep the autopoiesis of society going as there is always something to say about what has just been said. Even silence is rich in potential meaning. What counts is whether uttered information is \emph{accepted} as meaningful by observing systems.

Consequently, society is the \textit{observations of the observations of observers}, giving rise to an image of society as something comparable to organic complexity and contingency. As in life systems, no one social system can steer or determine the operations of another; instead, closed systems engage with one another only at the level of second-order observation. Every communicative event is therefore open to an unforeseeable range of possible second-order observations in its environment, which in turn can generate other environmental observations that in turn produce further responses.

Today's society is characterised for \citeauthor{luhmann1992communication} by \emph{functional differentiation}. Much of \cites{luhmann2012TOS1} extensive scholarly writing is concerned with tracing the evolution of social systems into higher degrees of complexity over time, which \citeauthor{luhmann1992communication} attributes to communicative differentiation. He distinguishes between three broad structures. \emph{Segmentary} societies are differentiated by the inclusion or exclusion of their members. \emph{Stratified} societies assign members positions in vertically hierarchical structures. In contrast, society since the development of modernity is characterised by the \emph{functional differentiation} of autonomous systems, non-hierarchically ordered, operating today on a planetary scale.

Broad examples of function systems are law, politics, economics, science, art, religion, education or philosophy. Each social system can be identified by its own binary ``coding'', i.e., core operative distinctions that are applied in that system's communication. The application of the code unfolds through the aid of ``programmes'' -- i.e., procedures, methods and structures of epistemic importance -- that serve to reduce the complexity of the environment and give each system its self-referential identity and functional value in the social systems as a whole. Each system can observe the others and communicate about them: legal decisions have economic effects, philosophical arguments have political implications, and so on. But they cannot enter into each other's communication directly.  

It may be objected at this point that \cites{luhmann2012TOS1} depiction of society as strictly differentiated autonomous systems is too rigid and formal to capture the polycontextual reality of social life, which is full of mixtures and hybrids, or that it negates some determining force such as power or economics. However, as systems theory's arguments apply to itself as much as to the rest of society, the theory designates itself as a sub-system in the social system of science. It insists that when we talk about social systems we can do so only by making use of another social system. 
In other words, the function of a system is defined by its environment and not by the system itself. The function of a social system is not found by studying the techniques and practices that constitute its internal structure. When considering XAI, similarly, the social function of a tool is not reducible to its computational operations but instead owes its meaningfulness to the observer. %

Second-order observation is another key characteristic of functional differentiation. Social systems are concerned with not only \emph{what} they observe of the world by distinguishing system from environment, but also with \emph{how} the world is observed through that distinction. Today, we are used to second-order observation. Communication is reflexively adapted to constant changes in its environment, and contingency is expected.

For instance, the legal system, which operates by communicating about the coded distinction between legal and illegal, constantly adapts its rules and procedures to process and stabilise changing events in other systems. Similarly, new scientific methods replace older ones, creating new programmes for determining the application of the scientific code of truth and false, as applied to cognate facts, in the face of new research. In politics, new channels for processing political inclusions or exclusions can be opened (or closed off) by political decisions. Individuals observe themselves as others see them through social media profiles. The contingency of communication, and society's self-awareness of this contingency, recursively increases the complexity of social systems, which become further differentiated and consequently generate further contingencies in a recursive process of self-differentiation.  

Science is, of course, an important social system for XAI. Observed functionally, it provides society with factual truths produced via second-order programmes that enable first-order observations to be verified. In order to be accepted as scientifically true, both the results \emph{and} the methodology of an observation must be painstakingly recorded and verified, including by a community of peer-observers, and falsification is accepted as a possibility. There are no absolute scientific truths independent of conditions of observation. The \emph{what} and \emph{how} of observation condition one another. XAI researchers must manage the oscillation between the factual output of ADM systems and their manner of observation; in a sense, this oscillation is the central problem. 

We can relate this point to the concept of ``link certainty'' developed by \citet{sullivan2022understanding}, %
which is crucial to understanding an ADM system, particularly when used to generate scientific or medical insights. It refers to ``the extent to which the model fails to be empirically supported and adequately linked to the target phenomena''. As ADM algorithms do not follow the scientific method when producing their models from data, the  meaningfulness of their outputs depends not on the transparency of the computational operations that determined how, technically, the ADM process arrived at a given output, but on the second-order question of the extent to which it works as a proxy for the same phenomena, understood scientifically. Link certainty therefore depends on the observational context. Its importance depends on what is at stake in the particular model's functional case. And this, too, is a second-order question.

Another example relevant to XAI is the legal system. For \citeauthor{luhmann1995social}, law has the function of maintaining normative expectations in the face of disappointment. An event that has caused disappointment -- i.e., ``that should not have happened'' -- is assessed according to the programmes of the legal system, which absorbs and proceduralises disputes. Only after ``due process'' can the legal status of a contested event -- the coding of legal and illegal -- be applied and accepted as legitimate by all parties. What the law requires of an explanation for a contested decision depends on the situation.

An explanation in itself is not a legal remedy, although it may form the basis for deciding whether or not a remedy is owed. Thus decisions reached by way of ADM systems, even if they arrive at the ``correct'' answer in a given case, may not satisfy the legal system's requirements for legal validity~\cite{deakinandmarkou}.
Recalling \cites{ackerman2003sociotechnicalgap} definition of the sociotechnical gap as the distinction between what we need socially and what we can do technically, it is possible that the normative expectation of understanding the reasons for a decision that affects legal rights (a social need) may lie beyond the technical capacity of machines~\cite{hildebrandt2018algorithmic,erasmus2021interpretability,deakinandmarkou}.

Hence researchers have warned against assuming that generating a successful \emph{understanding} of a particular decision is sufficient to \emph{legitimise} the use of an ADM system in contested adversarial contexts found in the legal system. In contrast to scientific research, which presupposes the alignment of interests between explainers and explainees, in adversarial legal scenarios the explainers and explainees likely have competing interests and each will be motivated to select explanations that suit their interests.  
As such, selecting an appropriate explainer cannot always be treated as a question of choosing the best -- ``most truthful'' -- option. Assuming otherwise, and building organisations on such assumptions, will ultimately produce trivial, partial, misleading or otherwise harmful effects by justifying illegitimate, unfair or discriminatory data-driven automation and decision-making~\cite{bordt2022post}. 

These examples illustrate how systems theory can incorporate a diverse range of sociotechnical literature. %
Indeed, all academic literature is part of the autopoiesis of society's self-observation. \cites{luhmann1992communication} elementary contribution is that assessing the meaning of the output of an explainer depends on the second-order observations of the system in question. At the same time, social systems are adaptive to changes in their environment through self-reference, evolving in response via contingent differentiation and adaptation. 
XAI tools should therefore aim to make available the information necessary for social systems to incorporate outputs of ADM systems according to their own social coding, which is internal to the differentiated operations of each system. This is not a purely technical matter, but requires extensive situational engagement with the problems that the particular ADM tool is intended to solve. 

Secondarily, this means it is unsurprising that a novel problem like XAI is multivariate. The term spans a  multiplicity of different problems for different systems, with science, philosophy, economics and law being obvious examples, each of which contains sub-systemic examples.

\section{Operational Closure and Structural Coupling\label{sec:closure_coupling}}

The \emph{closure} of social systems makes the theory an unusual way of thinking about society as human beings are excluded from society. Only communication communicates and communication is the systemic coupling of information, utterance and understanding. The ``observer'' is an abstraction, the necessary correlate of working with the primary distinction between system and environment. Who or what draws the distinction depends on the given situation. Observing the distinction requires a further distinction, which presupposes another observer, and so on. Hence the observer is always an abstraction -- a role that could be played by a single person at multiple levels, by an organisation or by multiple observers mediated via technological processing. The abstraction makes the theory flexible across contexts. 

The human being in the environment of communication is composed of the coupling of two operationally distinct autopoietic systems: the \emph{living} system of the body and the \emph{psychic} system of consciousness. The mind and body are coupled to society through the sensory organs as well as the linguistic and perceptual capacities of consciousness. Living, psychic and social systems are \emph{co-evolutionary}. They evolve independently of one another using their own specific operations, yet they depend upon, influence and constrain one another in ways that are contingent and unpredictable.

The closure of systems has consequences for ontological claims about the world. For instance, rather than a physical quantity, information is treated as \emph{a difference in the state of an observing system}. Information succeeds in changing the state of an observing system only when a meaningful difference is registered within the observing system; that is, if it creates a ``difference that makes a difference'' for the system~\cite{bateson1972difference}, rather than repeating what is already known.
As a result, ``information is different for everyone and always relative to a specific observer''~\cite{esposito2022artificial}. %
In \cites{luhmann1992communication} own words: ``[the] distinction between the information value of [the communication's] content and the reasons for which the content was uttered [is important because] the information is not self-understood but requires a separate decision [--] communication provides many possibilities for an accompanying perception''.%

Therefore, partners in a communicative interaction are in a position of \emph{double-contingency} with respect to one another. Thoughts and perceptions are internal to the organism or organisation and therefore are inaccessible to the other, meaning that neither party knows with certainty that their observation of an utterance yields the same understanding that the speaker intended it to have. Ego (self) and alter (the other) must each select their own informative aspects from their environments, which include each other, and must be able to recursively check their selections in order to secure functional understanding. Neither party is in control of the situation. Communication must solve the problems of communication using communication. 
Ego and alter have options for adapting their use of language and other perceptual media to the problem of mutual understanding while both \emph{knowing} that the other shares in the problem. 

Hence knowledge of what one does not know and the mutual ability to anticipate what the other may not know are crucial aspects of dealing with double-contingency. This cognitive attribute, which has evolved in complex organisms, is the catalyst for social evolution. In other words, social systems emerged as ways of stabilising communication between closed organisms, allowing the fundamentally improbable achievement of understanding to be reproduced with a higher degree of probability (but never with perfect certainty). Over time, the out-differentiation of communicating systems led to the emergence of world society. 

In this way, social systems theory is interested in the question of how communicating systems manage to reduce the complexity of their environments in order to remain relatively stable over time. A key concept in this respect is the linking of systems through \emph{structural couplings}, which are conceptual points at which systems can mutually resonate or ``irritate'' one another in a regularised and predictable manner. Structural couplings narrow the possibilities of each system's response to other systems and in so doing ``digitise analogue relations''. They reduce the uncertainty of the environment by providing specific interpretative grids for one system to observe another, allowing precise information to be formed within both systems; for instance, linking the social system of communication and the psychic systems of human beings as communication occurs~\cite{luhmann2012TOS1}. %

The concept of structural coupling can be observed in many phenomena but in the context of XAI it is perhaps best thought of as communicative media -- most prominently language -- that permit systems to connect tightly. Tightly coupled systems can generate, for example, mutually informative thoughts (in humans) and communications (in society), giving some predictability to otherwise contingent operations. To give other examples, a constitution is a mutual reference point for the interaction of the legal and political systems, which each communicate about it differently; a contract is both a legal and an economic artefact; and so on. 

Language, at the most general level, allows for complex human interactions. Our psychic systems cannot directly participate in communication -- and communication cannot ``think'' -- but communication could not continue without linguistic couplings to autopoietic conscious systems, which in turn depend on living human bodies. Autopoietic systems evolve in relation to one another yet independently of one another. This axiomatic position makes systems theory a useful tool for thinking about the integration of communication with ever-advancing computational systems. As \citet{esposito2022artificial} notes: ``no sharing of thoughts among [communication] participants becomes a great advantage when dealing with algorithms that do not think''. %

Technological systems, including computer systems, are not autopoietic. They are \emph{allopoietic}, meaning that they change not through internal operations but only when acted on externally~\cite{luhmann2018org}. For example, although learning algorithms display evolutionary properties, computers do not reproduce their operations spontaneously and they do not have operational independence from the social system of science that programmes, prompts, changes, controls and limits their operations. %
Technological systems are intrinsically coupled to society's communication, and indeed to many of the living systems of the body, and this has contingent consequences for the evolutionary yet independent systems of bodies, minds and society -- consequences that seem to be unfolding at an accelerating pace. But technology remains in the environment of society, a point we expand upon in view of communication media in Appendix~\ref{sec:communication_media}. %

Although not autopoietic, technical systems can nevertheless be operationally closed. Recall that \citet{luhmann1992communication} stipulates that direct information transfer between closed systems is impossible. %
Instead, each system is the environment of other systems and each must draw the distinction between system and environment for itself. 
Computers separate their information processing operations from understanding. They do not understand their operations, they simply carry them out and they do so invisibly to human perception. As \citet{luhmann2012TOS1} put it, digital media operate with a ``depth'' in excess of what is made visible on the surface as it is perceived. 

The screen is where uttered information is usually made observable by humans, and the graphical user interface is thus how computers are most commonly coupled to the social system, using language for their high-level operations that are conducted via binary processing. Clearly, not everything relevant to the user's understanding can or should be communicated on the screen. But as computers do not have self-perception, they \emph{do not know that they do not know} and cannot understand what has not been understood by the other system observing them. Therefore, unlike humans whose cognitive operations are coupled to society by language and self-perception, computers cannot innately learn to generate explanatory accounts of their own operations.
Hence as digital computers increase the scale of potential available information, as well as the processing capacity to analyse it, they also decrease the probability that what is selected as informative by the programme will be accepted as informative by other observers. Writing in the 1990s, \citet{luhmann2012TOS1} suggested that
``as ability increases, so does inability (as measured against it). The possibilities of arguing by accessing the invisible machine are clearly diminishing and the susceptibility to failure is increasing''. %

In this respect, the rise of predictive algorithms has a disruptive effect on the temporal structure of communication. Predictions output by machine learning models stem from patterns and associations found through algorithmic analyses of stored data, which are the accumulation of past observations. They do not depend on any prior understanding of those observations or patterns and associations that emerge in data processing. The achievements of ADM systems are potentially informative insights that arise from the fact that digitised data can be successfully \emph{analysed} without being \emph{understood} in advance. 

Whereas earlier attempts at achieving artificial intelligence envisaged programming computers to simulate intelligent living systems, machine learning models instead apply algorithmic procedures to large volumes of data at a scale that was simply impossible until recently~\cite{esposito2022artificial}. This is in large part thanks to the networked collection of data derived from digital communications that is the Internet. The results of machine learning systems are interesting because they cannot be anticipated or understood in advance and are not derived by deduction.
Algorithmic predictions about the future do not rely on generating averages, norms, ideal models or any other linguistic representation of people or the world~\cite{esposito2022artificial}. %

For systems theory, all technical information processing takes place in the \emph{environment of society}. The demand for explanations, as we have shown above, is always generated by observing social and psychic systems that need to contextualise new informative selections derived from the computer systems in their environment in order to understand their outputs at the second-order level and thus relate them to the ongoing autopoiesis of society. In other words, in order to accept new information in communication -- to understand it and therefore use it in further communication -- we need it to be rendered into a form that can mediate, or structurally couple, the linguistic and visual representation of information on the ``surface'' with a representation of what occurs in the ``depth'' of the algorithmic system. Therefore, we stress that generating an acceptable explanation is not the same as understanding the empirical digital operations of the algorithmic system in any precise scientific sense. As noted by \citet{esposito2022artificial}: ``algorithms as communication partners can be explainable without being understandable''. %

Note that the concept of operational closure resonates with the related cybernetic concept of \emph{merkwelt}. This term has proved influential in understanding closed biological and robotic systems, and has recently been productively applied by \citet{dazely2021levels} in theorising XAI. %
In brief, every closed cognitive system, whether living or robotic, produces its own internal construction of the world around it. %
\citet{dazely2021levels} argue that XAI tools should be approached as systems that have a cognitive relationship to the world, and accordingly they differentiate several systemic ``levels'' at which explainers should be designed. These are: %
\begin{description}
   \item[zero-order explanation] by which the explainer reacts automatically to a data-driven decision-maker in order to analyse its output;%
  \item[first-order explanation] processes the analysis in relation to the intention, goal, function or aim that the explainer system is expected to address; 
  \item[second-order explanation] incorporates the explainer agent's predictive analysis of the expectations of the cognitive agents in its environment (human or otherwise), aiming to successfully anticipate their anticipations and thus address their second-order observations of the explanation; and
  \item[n\textsuperscript{th}-order explanation] concerns higher levels of abstraction, involving the norms of the ``culture'' in which the agent is to be deployed.
\end{description}
\citet{dazely2021levels} note that the agent should be capable of measuring the explainees' ``quiesence'', that is, be capable of cognitively responding to signals that indicate whether the explanation has been successfully accepted. We can understand this \emph{Broad-XAI} approach as a design paradigm for artificial cognitive systems that can operate with society as their environment.

However, even if evolutionary approaches to developing such cognitive agents are employed, we argue that this remains a technical question of design rather than an evolutionary achievement of autopoietic self-organisation. The goal is not to bring computers into communication but to aid the emergence of what \citet{esposito2022artificial} terms ``artificial communication''. 

\section{Towards Explanatory Media\label{sec:communication_media}}

Based on the observations outlined in Appendix~\ref{sec:closure_coupling}, %
the challenge of XAI is in developing ways to successfully reduce the contingencies of understanding different ADM systems in relation to different social systems.
The technical difficulties go beyond the algorithmic selection of \emph{information} and \emph{utterance} described in Section~\ref{sec:selections}, and, among others, include social systemic problems such as systematically anticipating different users with different needs, reporting on link uncertainty between ADM models and scientific models, and incorporating regulatory rules and requirements. %
We suggest here that the multiplicity of concurrent attempts to design, develop, test and deploy XAI tools represents an evolutionary passage in the history of communication media and we hypothesise that eventual explainability could be regarded as an emergent sub-system of scientific communication. %

Systems theory has a complex approach to media. Media cannot be simply treated as hardware that passively performs the transmission, storage and processing of information because the theory is primarily concerned with distinctions that generate meaningful observations by indicating one side of a form and excluding the other. Every observation is constituted by a form, which means that observation always depends on a medium that can be divided and marked by a form. A form is thus regarded as a ``tight coupling'' of ``loosely-coupled'' elements of a medial substratum~\cite{luhmann2012TOS1}. %
The form has a marked side and an unmarked side, corresponding to the distinction between system and environment that separates what is indicated from what is excluded. As explained in Appendix~\ref{sec:communication_systems}, either side can be picked up and utilised in further second-order communication. This implies that when we designate something as a medium, we must use a form, which in turn presupposes another medium that can be marked. Hence there is no material substrate upon which communication ultimately depends.

To give a trivial example, a sentence is a form that can be observed in the medium of words, while words are forms in the medium of letters. Yet all text can be observed as forms marked on the medium of paper or indeed on any physical object capable of bearing them -- one can write one's name on the sand. Thus the self-referential, autological structure of systems theory manifests itself again: without media there can be no forms, but without forms there is no way to indicate media. Forms can be deployed in different media, while media make themselves available to different forms, other possible selections and indications~\cite{luhmann2012TOS1}. %
 
There are two ways in which the distinction between media and form is relevant to XAI. First, it allows us to situate algorithmic media in the history of \emph{dissemination media} through a brief illustration of how society has been transformed by advances in technological media; we compare oral, written and digital media to this end (Appendix~\ref{sec:communication_media:dissemination}). Second, XAI involves engagement with \emph{symbolically generalised communication media}, or \emph{success media}, which are the semantic media that provide contextual shared reference points in social systems, thus allowing participants to relate new information to existing understandings and making it more likely that communication will succeed (Appendix~\ref{sec:communication_media:symbollical}). %

\subsection{Dissemination Media\label{sec:communication_media:dissemination}}

At the most elementary level, pre-lingual perception permits understanding to arise amongst humans, and also with animals, as long as the perception of perception is possible. Once mutual perception occurs, it can form the basis for sharing meaningful signs and gestures~\cite{luhmann2012TOS1}. %
In oral communication, meaning arises in the form of words distinguished from the medium of the voice. At the level of voice or gesture, the tripartite structure of \emph{information}, \emph{utterance} and \emph{understanding} coincide in time and space. Alter speaks, ego listens and acceptance or rejection follows almost immediately; the process is fuelled by (mis)understandings that prompt the need for clarifications, disagreements and so on.

We can become aware of the distinction between medium and form -- for example, between voice and sound, or more complex differentiation such as singing and voice or questioning and voice (i.e., vocalised forms) -- if we observe it in a further moment of second-order observation, suspending the issue of what was said and instead considering how it was said. Language stabilises these possibilities, making oral societies possible, and allows second-order meta-communication to emerge as it is then possible to indicate in language that one is communicating in language. 

Writing did not begin as a mechanism for communicating over space and time -- as that would have presupposed readers -- but first emerged in religious rituals and as a mechanism for managing lists of items in economic transactions~\cite{luhmann2012TOS1}. %
Only over many centuries did it engender different forms of communication, forms that also depended on the development of suitable dissemination media, including paper, print, postal networks as well as electrical and electronic networks. Writing allowed communication across space and time, eventually via technical dissemination media such as books, newspapers, letters, fax and e-mail.

But what counts is not the mere fact of transmission and storage of information but the proliferation of meaningful forms of communication and the momentous changes in culture that followed, including transformations in human understanding of time~\cite{luhmann1992communication}.  %
Writing disrupted the simultaneity of communication because understanding is separated from the moment of selection of information and utterance. A reader can encounter a text hours, days or centuries after it was written and this possibility, along with the contingency of the reader's understanding, is in turn anticipated by the author. This prompted and enabled the emergence of institutions and organisations that relied on the capacity of writing to carry forward information in time.

Text thereby allowed society to reflexively incorporate a new awareness of its own contingency in social, temporal and material dimensions. Observers have irreducibly different points of observation -- a realisation that has to be accommodated in communication. In the era of functional differentiation, unlike structures of society that bound together politics, religion and legality in strict hierarchical structures, the availability of stored written information and the multiplicity of possible interpretations ultimately allow individuals to be freed of any requirements to learn and identify with particular knowledge.

Instead, what counts is the ability to retrieve information for use in communication -- a question of access rather than learnt knowledge -- based on indexing the relevant data, with the library catalogue and the file registry being prime examples. Indexed data, whether stored in organisations, libraries, universities or personal diaries, allowed for a stable form of social memory and thus enabled the forgetting of information in ongoing communication, freeing up cognitive capacity for new information. One need not know everything to deal with the complexity of modern society, but to navigate complex systems, one must learn to access information.   

Digital media sever the links between all three selections. With computers, which can recall and combine data from past information to generate new information, the elements of \emph{understanding}, \emph{information} and \emph{utterance} are all potentially mutable in time and space~\cite{luhmann2012TOS1}. %
In this respect, the recombination work performed by learning algorithms has a particularly disruptive effect on the temporal structures of communication. For example, predictions output by machine learning models are derived from new analyses of previously communicated information in the form of stored data. As we discussed in Appendix~\ref{sec:closure_coupling}, this does not depend on prior understanding of that information or of its previous associations. %

The informative aspects of ADM algorithms is derived from their second-order data processing work. Hence society's past, in the form of stored data, comes to inflect upon the future decisions of society via the new insights that ADM systems produce. We refer again to \cites{esposito2022artificial} succinct point that the utility of ADM tools lies precisely in their separation of \emph{information} and \emph{understanding}. This technical achievement leaves society with the problem of retrospectively understanding what has emerged from the oracle of the algorithm. Unlike societies that believed in divine revelations, second-order observation demands explanations. %
But can explanations themselves become a generalised form and might this be a cause for concern? To address this, we turn to the notion of ``success media''.

\subsection{Symbolically Generalised Communication Media\label{sec:communication_media:symbollical}}

Framed as the contingent distinction between \emph{information} and \emph{utterance} that must be simultaneously achieved by two or more interlocutors, successful communication is highly improbable. This begs the question of how society has achieved stability in its operations. While we have so far been discussing how systems process new information, it is equally important to consider the redundancy that allows new information to be related to existing understandings. This stability is required as systems selectively generate new information about their environments.  %

One answer is that systems make use of symbolic ``success media'', or \emph{symbolically generalised communication media}, which enhance the acceptability of new understandings. If new information can be processed in a form that makes use of these media, they will be more readily accepted by the system. For example, the legal system's symbolic success medium is justice, a symbol that is generally available to describe situations (whether just or unjust) in legal communication. ``Justice'' is an abstraction, a symbol with no single definition. It is an inherently contested concept. Precisely due to the lack of a concretely agreed definition of justice it is available as a symbol that can take different forms in different situations. In general -- and certainly with regard to its organisational elements such as courts, institutions and officials -- the modern legal system regards justice as a self-generated procedural question. Debates persist about the requirements of justice, and one can still claim in a religious modality that justice is part of ``nature'', but the successful acceptance of legal communication no longer depends on the old forms of ``natural justice'' that structured the legal system prior to functional differentiation. 

If someone has a legal complaint -- as many people do -- they are more likely to receive any positive response in the legal system if they adopt the forms and protocols of the institutions of justice. This is why lawyers, with their training and hefty fees, are socially important and why the substantive outcome of a case is likely to be accepted as valid provided the correct procedures have been followed. This is the case despite the contingency of positive law. Today, the legislature or executive might decide to change the rules so that what was legal yesterday becomes illegal tomorrow. Hence ``justice'' is, as \citet{luhmann2012TOS1} put it, a formula for absorbing the contingency of legal decisions. %

Other examples of symbolically generalised media that absorb the contingency of decisions in communication include love in the system of the family, money and property in the economic system, power in the political system, and truth in the sciences. 
Such media cannot be precisely defined. They are observable via the specific forms they take in specific social situations, for instance, a kiss, a court hearing, a payment, an order or an experimental result. Media thereby stabilise communication and give substance to social memory and structural couplings, allowing greater complexity to emerge. Thanks to their stabilising role, greater varieties of differentiation can emerge and stabilise over time. 

In view of these observations, %
we suggest that XAI might be framed as a project of developing technically-generated symbolic success media. The presumption is that information produced by ADM algorithms will be more acceptable if accompanied by good explanations. There are, as we have indicated in Section~\ref{sec:prelim}, many ways of generating an explanation and of trying to make it understandable. 
Insofar as this can be achieved technically, it can be observed as the question of producing ``XAI'' as a new medium that, in different forms, increases the likelihood of both an ADM system and its explainer being integrated into society as a component of artificial communication. %

If this is the case, then the project is a sociotechnical hybridisation of technical design and communicative evolution. This points to the risks of XAI to existing success media. If algorithmic explainers come to have their own media of success, they could enable algorithmic decision-making to displace some forms of other success media. In other words, the decisions of automated systems could come to take precedence in society over just legal processes or scientific truths if the medium of explanation successfully takes over the functions assigned to justice and truth respectively. 

The question of whether explanations are stabilised and deployed in society does not simply turn on whether or not XAI develops independent success media, but also on whether or not such media can solve social problems that are currently solved with other media. For instance, might an explanation substitute for a legal process? The risk that algorithmic tools might displace the function of legal or political communication in making decisions in society has been highlighted, among others, by \citet{hildebrandt2018algorithmic}.
Concretely, such developments will depend on the capacity of decision-making structures in society to incorporate and operationalise insights produced by ADM systems as fresh information. Hence in Appendix~\ref{sec:organisations} we turn to consider the role of organisations.

\section{Organisations and Accountability\label{sec:organisations}}%

This paper is predominantly concerned with the %
abstract aspects of communication, thus far only superficially relating it to interpersonal interactions. %
But to the extent that these themes are activated in social practice, it is necessary to consider how communication is processed in organisations, %
of which interpersonal interactions are a distinct from. %
Whereas for the latter interactions activate couplings between psychic and social systems under conditions of double-contingency, the former process and perceive communication differently, as anyone who has worked in an organisation may have noticed. %

There is a gap between what organisations communicate and what their members think.
Critical analysis of the design and deployment of XAI in society requires awareness of this distinction. %
If ``explainability'' becomes a medium for securing the acceptability of decisions output by ADM systems, as we suggest in Appendix~\ref{sec:communication_media}, then the forms that explanations produced by XAI tools take may be organisational -- e.g., modes of hierarchical authorisation, accreditation schemes, taxonomies, regulatory standards, insurance schemes or review boards -- rather than individual. %
In short, the uncertainty that individual users may have about the meaning of an ADM output may be absorbed within organisations by second-order procedures.
Arguably, this is already occurring in society as companies rush to make systems based on artificial intelligence algorithms available to the market without understanding their operations. %

Organisations can be observed as closed autopoietic systems that connect and stabilise \emph{decisions}, marking time as new decisions are produced in response to past internal decisions as well as to environmental changes. They learn, developing techniques for setting goals, generating information, deleting or retaining memory, distributing decision-making powers, conducting reviews, anticipating risks, carrying out plans, and reviewing the relationship of the organisation to its operational environment.

Operationally, decisions are produced by programmes and procedures that allow organisations to impute causality to their own operations. Through such decision programmes responsibility, transparency and accountability are performed as technical operations. 
In this way organisational forms can manage the coupling of differentiated communication by having departments or members tasked with financial, legal, operational and mass media (PR) relations, among many others. This makes organisations and their members suitable targets for legal regulatory frameworks. Developing decision-making responsibility and accountability procedures within organisations is a second-order process by which they both self-regulate and adapt to external regulatory requirements communicated in the legal system. %

As such, addressing these aspects of accountability begs second-order questions such as ``Accountable to whom?'' and ``Accountable on what terms?'' The critical viewpoint presupposes an observer who is usually another organisational construct in the environment of the system such as an independent regulator or a political authority. The accountability requirements of one observing system will be different to the requirements of others such as users, investigators, lawyers, insurers and the public, or more precisely the mass media that stand in for the public~\cite{dazely2021levels}. %
Organisational communication, in other words, enables communication to adapt and absorb the uncertainty of navigating a complex environment~\cite{luhmann2018org}. %
Organisational responsibility involves the tailoring of accountability procedures in ways that are irreducible to the understanding of a situation in the psychic systems of individual members.

The well-known figure of the human decision-maker -- or the ``human-in-the-loop'' paradigm that is often referenced as a means of controlling algorithmic decision-making -- is frequently offered as a focal point for representing organisational responsibility and accountability procedures. %
Yet this hypothesised role for human consciousness seems to miss the distinction between psychic and organisational systems in communication.  %
An organisation manages time and distributes responsibility in ways that actively avoid individual decision-makers. 
An accountable individual may or may not have been consciously responsible for the decisions attributed to them or the consequences that follow~\cite{luhmann2018org}. %

Scholars critical of shallow forms of accountability discourse have recognised these problems, leading to calls to build regulatory norms into the architecture of ADM systems rather than relying solely on post-hoc accountability procedures and XAI tools~\cite{cobbe2021reviewable}.
We agree with the broadly stated goals but note that such a project also spans organisational processes through which design standards can be recursively generated and tailored for specific sociotechnical situations~\cite{sileno2018role}. %
XAI alone cannot solve problems of legal and political accountability. Attention must be paid to the deployment of algorithmic systems within the larger sociotechnical assemblages that they co-constitute~\cite{cobbe2021reviewable}.

Explainability is an important element of such assemblages but it is not the same as accountability even if an explainer tool were deployed successfully on its own terms. Explanations, by appearing to produce accountability, can serve to engender authority, arrogating power to the owners and users of the ADM algorithms at the expense of others. %
Na{\"i}ve approaches to the social implications of XAI risk perpetuating a false sense of security in adopting automated systems without regard for the harms that may be caused~\cite{aradaublanke2022algorithmic,green2022flawsofpolicies}. %
In short, critiques and observations from social sciences emphasise that the \emph{context} in which an explainee seeks and receives an explanation informs \emph{understanding} as much as the \emph{explanatory information} that is selected and presented; %
here, we argue that this context must include organisational forms. 
Yet critique will always arise if only because the sociotechnical gap between what is technically possible and what is socially demanded recurs as second-order observation. 

A broader discussion of organisational accountability is beyond the scope of this paper. %
The key point is that taking a systems theoretic approach to XAI indicates that it is important to include assessments of the organisational forms in which an explainer is intended to operate. %
Thus focusing exclusively on the psychic understanding and uncertainty of \emph{individual} users is largely insufficient. %
These aspects are of course necessary for successful integration of XAI insofar as individual users gain an understanding of ADM systems -- %
after all no organisational knowledge can emerge unless someone is in the know.
But excluding organisational models altogether risks failing to anticipate the ways in which organisations absorb and actively operationalise uncertainty.   

\end{document}